\documentclass[runningheads]{llncs}
\usepackage{graphicx}
\usepackage{xcolor}
\usepackage{tikz}
\usepackage{hyperref}
\usepackage{wrapfig}
\usepackage{listings}
\usepackage{subcaption}
\usepackage{calc}

\usepackage{amsmath,amssymb}
\usepackage{mathrsfs}
\usepackage{physics}

\usetikzlibrary{arrows,positioning,automata,circuits.logic.US,circuits.logic.IEC}

\usepackage{todonotes}

\begin{document}
\title{\textsc{MoonLight}: A Lightweight Tool for Monitoring Spatio-Temporal Properties \thanks{This research has been partially supported by the Austrian FWF projects ZK-35 and W1255-N23, by the Italian PRIN project ``SEDUCE'' n. 2017TWRCNB and by the Italian PRIN project ``IT-MaTTerS'' n, 2017FTXR7S.}}
\author{E. Bartocci\inst{1} \and
L. Bortolussi\inst{3} \and
M. Loreti\inst{4} \and 
L. Nenzi\inst{1,2,3}  \and 
S. Silvetti\inst{5}
}
%
%
\institute{TU Wien, Austria \and 
\email{laura.nenzi@tuwien.ac.at} \and
DMG, University of Trieste, Trieste, Italy \and 
               University of Camerino, Camerino, Italy \and 
               Esteco S.p.A., Trieste, Italy }

\maketitle              
\vspace{-2ex}
\begin{abstract}
We present \textsc{MoonLight}, a tool for monitoring temporal and spatio-temporal 
properties of mobile and spatially distributed cyber-physical systems (CPS). In the proposed framework, space is represented as a weighted graph, describing the topological configurations in which the single 
CPS entities (nodes of the graph) are arranged.
Both nodes and edges have attributes modelling physical and logical quantities that can change in time.
\textsc{MoonLight} is implemented in Java and supports the monitoring of Spatio-Temporal Reach and Escape Logic (STREL) introduced in~\cite{Bartocci17memocode}. 
\textsc{MoonLight} can be used as a standalone command line tool, as a Java API, or via \textsc{Matlab} \texttrademark\enskip interface.
We provide here some examples using the \textsc{Matlab} \texttrademark\enskip interface and we evaluate the tool  performance also by comparing with other tools specialized in monitoring only temporal properties.



\end{abstract}

\section{Introduction}
\label{sec:intro}

%
%

Cyber-physical systems~\cite{Ratasich2019} (CPS) are a widespread class of technological artefacts that include
contact tracing devices, self-driving cars,  mobile ad-hoc sensor networks and
smart cities.  
CPS are controlled by a computational device and interact within the physical space. As such, they are described by discrete states, controlling actuators, and continuous quantities, measured by sensors, which can both change in time. CPS are arranged in spatial configurations that can be static or dynamic. Their network connectivity can typically change in time. 

A fundamental task in engineering CPS is monitoring their behaviors, specified in a suitable formal language, such as Signal Temporal Logic (STL)~\cite{mn13,MalerN04}. Monitoring can be performed on a deployed system or on simulations of a model, as typically done in the design phase to test different initial conditions, parameters and inputs~\cite{chapter5,BartocciFFR18}. 
Monitoring a trace returns either a Boolean value, witnessing whether the requirement is satisfied or not, or a quantitative value, for instance a real-value indicating how much the specification is satisfied or violated according to a chosen notion of distance~\cite{FainekosP09,JaksicBGN18,BartocciBNR18,JaksicBGNN18,filtering}.

Current tools available for monitoring formal specifications are restricted to temporal properties, mostly ignoring the spatial dimension of CPS~\cite{chapter5,BartocciFFR18}.

\noindent{\emph{Our contribution}}.
We present \textsc{MoonLight}, a lightweight 
tool for monitoring temporal and spatio-temporal
properties of 
spatially distributed CPS, which can move in space and change their connectivity (mobile CPS). 
\textsc{MoonLight} is implemented in Java and supports monitoring of Spatio-Temporal Reach and Escape Logic (STREL), a spatio-temporal specification language 
introduced in~\cite{Bartocci17memocode}. 
STREL extends STL~\cite{mn13,MalerN04} 
with two main spatial operators \emph{reach} and \emph{escape} 
from which is possible to derive many other spatial operators (e.g., \emph{everywhere}, \emph{somewhere} and \emph{surround}). 
Our implementation is available at:
{\centering
\href{https://github.com/Quanticol/MoonLight}{https://github.com/MoonLightSuite/MoonLight}.
}
\textsc{MoonLight} can be used: as a standalone command line tool, as a Java API, or via \textsc{Matlab} \texttrademark\enskip interface.
In this paper, we describe the usage of \textsc{MoonLight} via \textsc{Matlab} \texttrademark \enskip interface because several CPS models and analysis tools are available for this framework.  We refer to the documentation for the other usage possibilities.



\textsc{MoonLight} takes as input a STREL formula
and a spatio-temporal trajectory. 
Space is represented as a weighted graph, describing the topological 
configurations in which the CPS entities are arranged. Nodes represent single entities. Both nodes and edges have attributes modelling physical and logical quantities that can change in time. Therefore, a spatio-temporal signal, in the most general case, is described by a sequence of such weighted graphs, allowing both spatial arrangement and attributes to change in time. 
\textsc{MoonLight} monitors such a  sequence of graphs with respect to
a STREL formula, returning a Boolean or a quantitative verdict, according to the semantic
rules of~\cite{Bartocci17memocode}.




  \vspace{-3ex}
\subsubsection{Related work.}
Monitoring tools for CPS are generally agnostic of the spatial configuration 
of the entities such as sensors and computational units.
They are limited to monitor  temporal specifications over 
time series of data. Examples are S-Taliro~\cite{staliro} for Metric Temporal Logic (MTL)~\cite{Koymans90},
R2U2~\cite{MoosbruggerRS17} for Mission Linear Temporal Logic (MLTL)~\cite{MoosbruggerRS17},
AMT~\cite{NickovicLMFU18} and Breach~\cite{breach} for
Signal Temporal Logic (STL)~\cite{mn13,MalerN04},
TeSSLa~\cite{LeuckerSS0S18} and RTLola~\cite{BaumeisterFST19} for
temporal stream-based specification languages and
Montre~\cite{Ulus17} for 
Timed Regular Expressions (TRE)~\cite{AsarinCM02}.  
However, temporal specification languages are not always expressive enough to capture the rich and complex spatio-temporal patterns that CPS display. 
For this reason, many researchers have extended temporal specification languages such as STL to express also
spatial requirements.  Examples include Spatial-Temporal Logic (SpaTeL)~\cite{spatel}, the Signal 
Spatio-Temporal Logic (SSTL)~\cite{NenziBCLM18}, the 
Spatial Aggregation Signal Temporal Logic
(SaSTL)~\cite{Ma20} and 
STREL~\cite{Bartocci17memocode}. 
Despite many developed prototypes
built more for demonstration purposes rather than becoming usable tools, 
we are aware only about \textsc{jSSTL}~\cite{NenziBL16} as offline monitoring
tool for spatio-temporal properties. \textsc{jSSTL}~\cite{NenziBL16} supports SSTL and operates over a \emph{static} topological space, 
while the tool proposed in this paper can also monitor
 \emph{dynamical locations}, such as in mobile wireless sensor networks.



  \vspace{-2ex}
\section{MoonLight in a nutshell}

\vspace{-2ex}
The main component of \textsc{MoonLight} is its Java Application Programming Interface (API): a set of specialized classes and interfaces to manage \emph{data domains} and \emph{signals}, to represent \emph{spatial models} that can evolve in time, to \emph{monitor temporal and spatio-temporal properties} and to manage input/output to/from \emph{generic data sources}. 
Moreover, it also contains a \emph{compiler} that
generates the necessary Java classes for monitoring from a 
\textsc{MoonLight} script.  The latter are built and dynamically loaded to enable the monitoring of the specified properties.

\textsc{MoonLight} provides also an interface that enables the integration of its monitoring features in \textsc{Matlab}\texttrademark.
%
We now first introduce a simple running example to guide the reader on how to monitor spatial-temporal properties.
Then, we present a \textsc{MoonLight} script together with a gentle introduction of formulas semantics. 
Finally, we show how \textsc{MoonLight} can be used in the  \textsc{Matlab}\texttrademark{} environment.

\vspace{-3ex}

\subsection{Running Example}
\vspace{-1ex}
A running example is used to describe the behaviour of our tool. We consider a wireless ad-hoc sensor network~\cite{Akyildiz2002} consisting  of three different types of mobile device:  
\emph{coordinator}, \emph{router}, \emph{end-device}. 
For each network, there is only one  coordinator that 
is responsible to initialize the network.  
The routers are responsible for passing data on from other devices and they establish a mesh network of intermediate devices to reach more distant ones.
The end-devices can only communicate with router or coordinator, and they cannot relay data from other devices.
%
%
We assume that all the devices are battery-powered and they are equipped with sensors enabling them to measure and collect data from the environment (e.g. pollution, temperature, etc.). 

\begin{figure}
	\includegraphics[width=11cm]{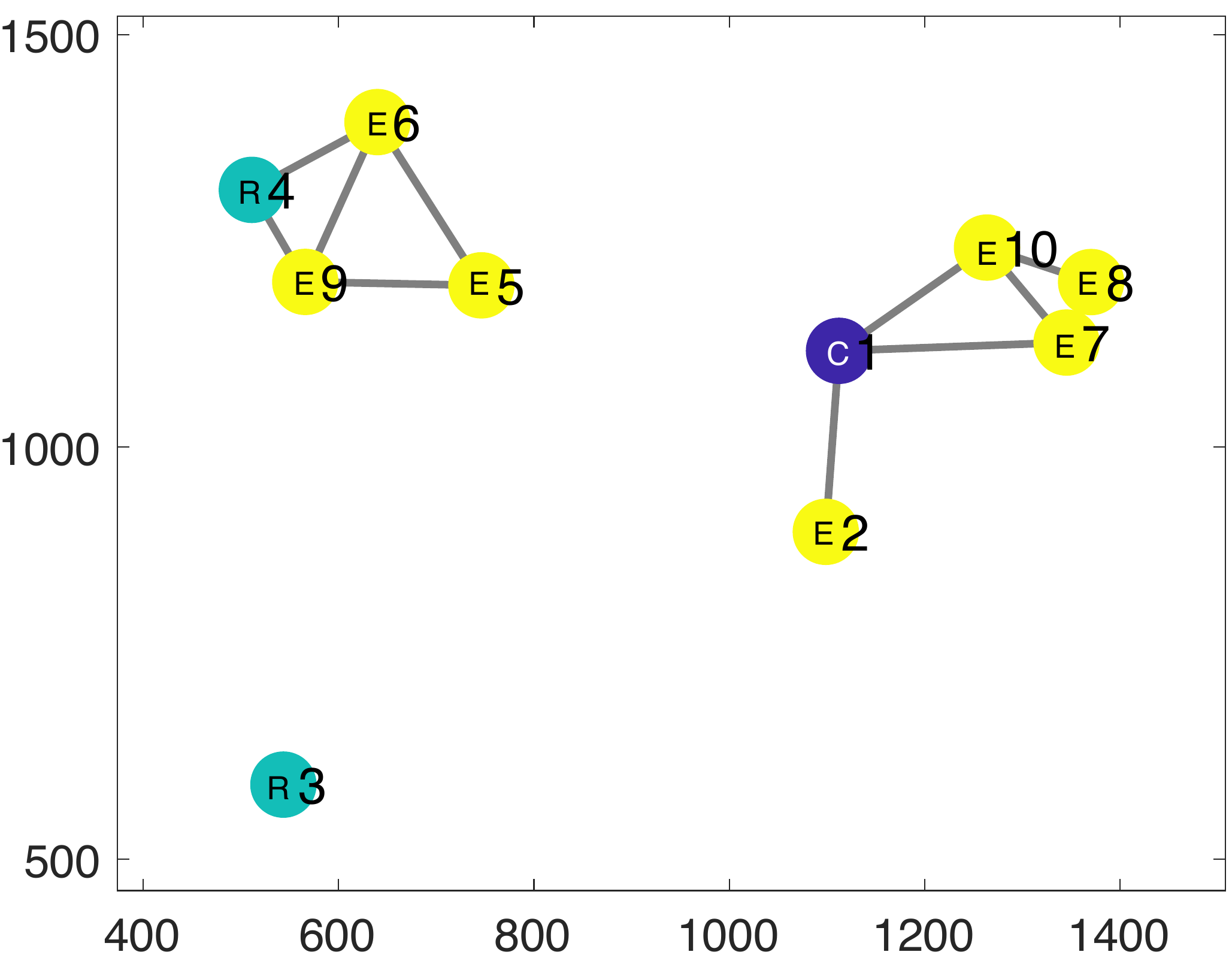}
	\caption{ Sensor network with 1 {\color{violet!100!black}coordinator} (C, in violet), 2 {\color{cyan!80!yellow!90!}routers} (R, in cyan) and 7 {\color{yellow!80!black}end devices} (E, in yellow).}
	\label{fig:graph}
\end{figure}

Fig.~\ref{fig:graph} illustrates a network with 10 nodes (1 {\color{violet!90!black}coordinator} (C) in violet, 2 {\color{cyan!80!yellow!90!}routers} (R) in cyan and 7 {\color{yellow!80!black}end devices} (E) in yellow). The nodes are distributed in an Euclidean space, i.e. axis represent their coordinates in the space. The edges represent the connectivity graph of the network, expressing the fact that two devices can directly \emph{interact} (i.e. they are within their communication range).

\begin{figure}
\hspace{-1ex}
	\includegraphics[width=6.1cm]{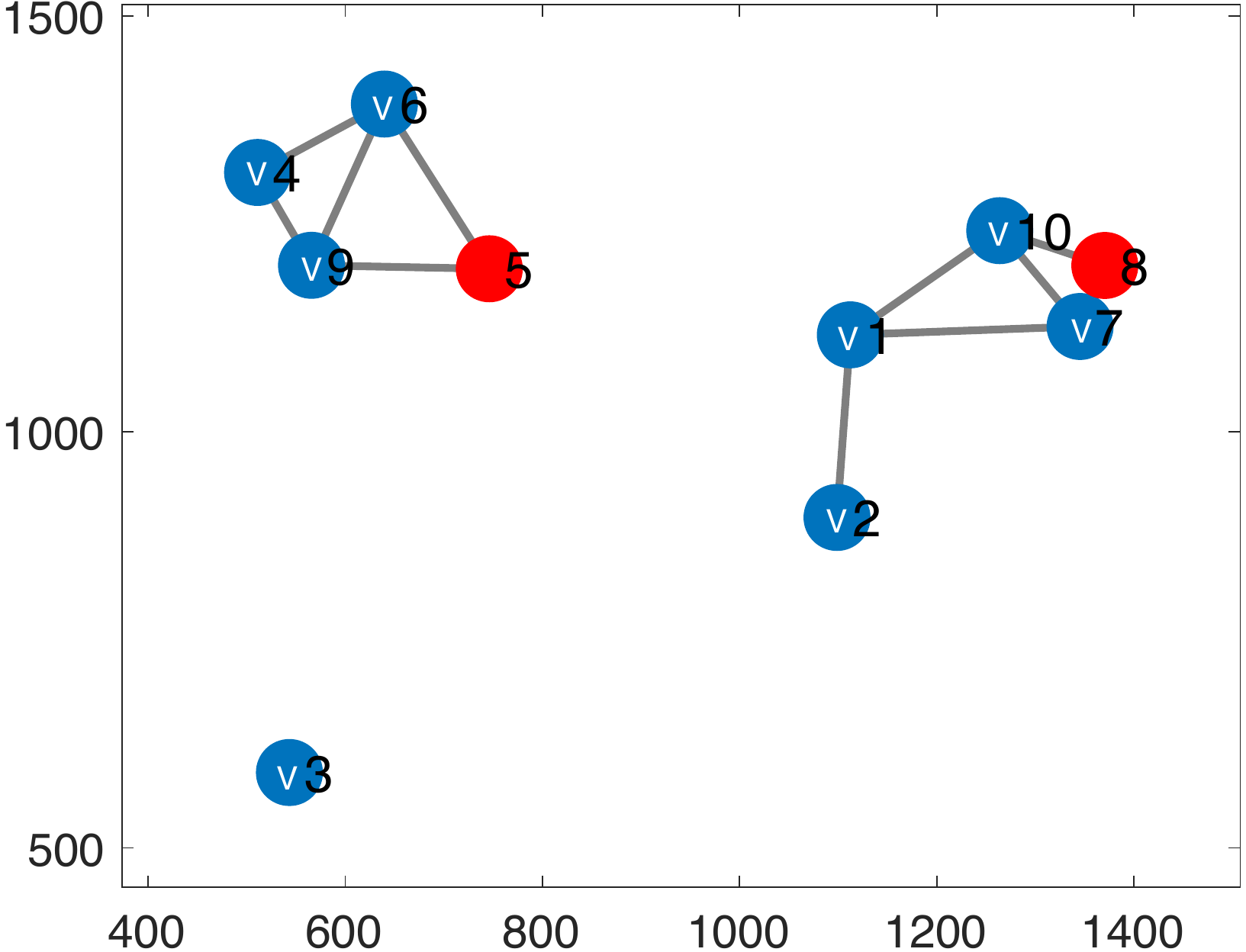}
	\includegraphics[width=6.1cm]{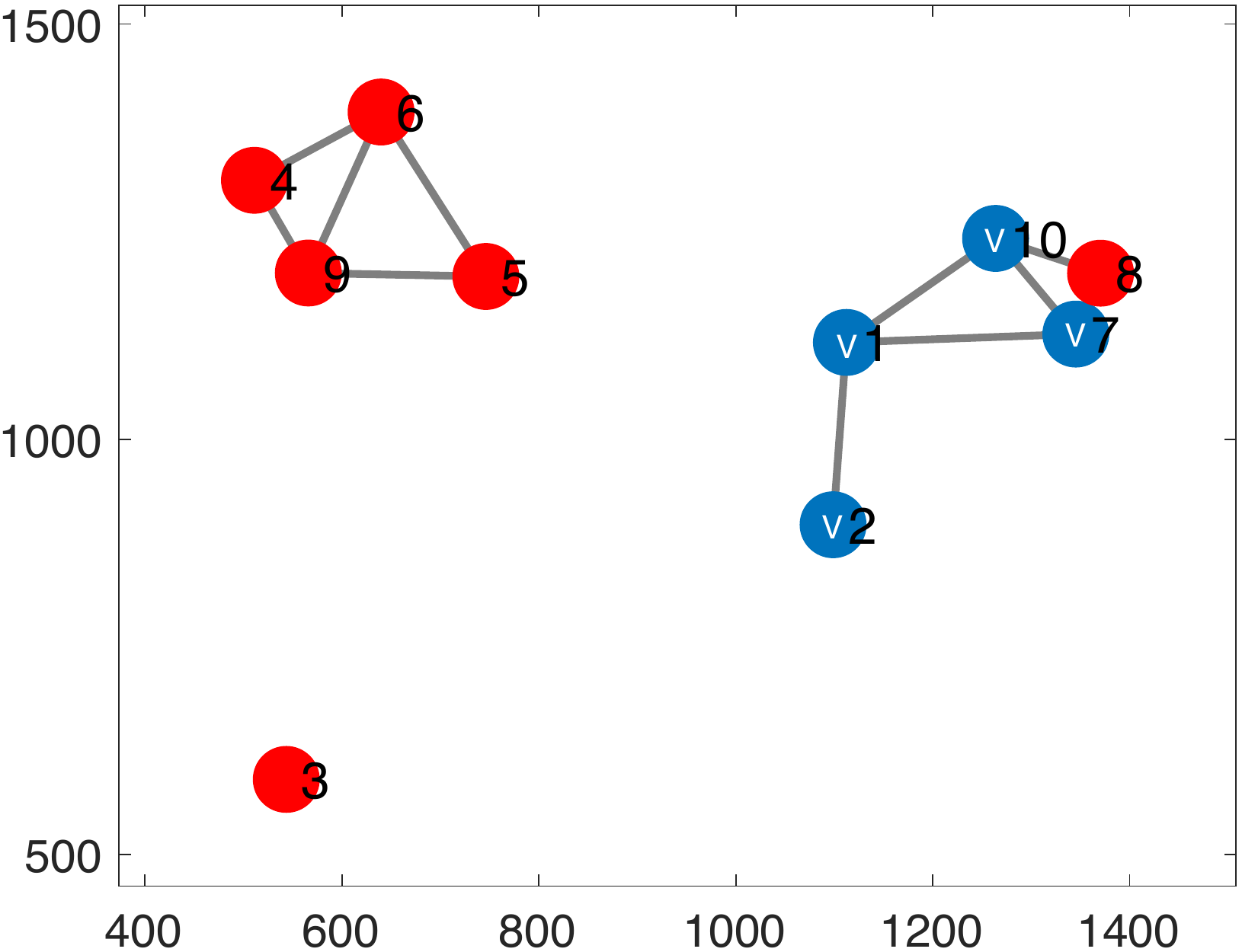}
	\caption{(\textbf{left}) Boolean satisfaction of formula \lstinline|P1|,  {(\color{blue} blue nodes} (V) satisfy the formula), {\color{red}red nodes} do not satisfy the formula.(\textbf{right}) Boolean satisfaction of formula \lstinline|P4|, {(\color{blue} blue nodes} (V) satisfy the formula), {\color{red}red nodes} do not satisfy the formula.}
	\label{fig:spatialEval}
\end{figure}


In \textsc{MoonLight}, the space is modelled as a graph,
where each node represents a location containing mixed-analog signals while each edge represents a topological relation. 
Edges are labelled with one or more quantitative attributes describing additional information about the spatial structure.  
In our example, the sensor network is our graph, each device represents a node/location of the network and contains three signals evolving in time: the type of node (coordinator, router, end-device), the level of battery, and the temperature. 
The edges are labelled with both their Euclidean distance and with the integer value 1. This last value is used to compute the hop (shortest path) count between two nodes, that is the number of intermediate network nodes through which data must pass between source node and target one. 

\textsc{Moonlight} evaluates properties specified in the linear-time spatio-temporal logic STREL over spatio-temporal signals, i.e. functions mapping each node and time instant into a vector of values, describing the internal state of each location. 
In the following, we show how to use the \textsc{MoonLight} scripting language to specify spatio-temporal properties and how to monitor them. 

\subsection{\textsc{MoonLight} Script}
\label{sec:moonlight_script}

A monitor in \textsc{MoonLight} is specified via a \textsc{MoonLight} script. Fig.~\ref{fig:script1} reportes an example that specifies the necessary information to instrument a monitor for our sensor network. 

\begin{figure}
\begin{lstlisting}
signal { int nodeType;  real battery;  real temperature; } 
space { edges { int hop; real dist; }  } 
domain boolean;
formula atom = (nodeType==3);
formula P1 = atom reach(hop)[0, 1]{(nodeType==1)|(nodeType==2)};
formula Ppar(int k) = atom reach(hop)[0, k] (nodeType== 1);
\end{lstlisting}
    

    \caption{Example of Moonlight monitor script specification, corresponding to the sensorNetMonitorScript.mls of line 1 in Fig. \ref{fig:script2}. }
    \label{fig:script1}
\end{figure}

The script (line 1) starts with the definition of the domains of input signal. 
We recall that STREL is interpreted over spatio-temporal signals. In our scenario, these values are: the type of node, the temperature level and  the battery level. As domains, the node type is represented by an integer (\lstinline|int|) and the battery and temperature by real values (\lstinline|real|).
Spatial structures in STREL can change over time.  This enables the modelling of the mobile network of sensors as in our example by updating edges and edge labels. 
The edges can have more than one label with different domains. These are specified in line 2 of our example.  In this case we have two labels:  \lstinline|hop| having type \lstinline|int| domain, and \lstinline|dist| with type \lstinline|real|. 
Note that, if one is only interested in temporal properties, this part is omitted in the script.

\textsc{MoonLight}, like STREL, supports different semantics for monitoring. A user can specify the desired one by indicating the specific \emph{monitoring domain} (see line 3).
Currently, \textsc{MoonLight} supports qualitative (\lstinline|boolean|) and quantitative (\lstinline|minmax|) semantics of STREL.

After this initial declaration, the script contains the list of formulas that can be monitored (see lines 4-6). Formulas can have parameters that are instantiated when monitoring is performed. A formula can be used within another formula. 

The syntax of STREL is summarized in Fig.~\ref{fig:syntax}. 
The atomic expression consists of \emph{Boolean expressions} on signal variables like, for instance, \lstinline|(battery  > 0.5)| or \lstinline|(nodeType == 2)|. 
As expected, the interpretation of atomic formulas depends on the domain of the monitoring.

Formulas are built by using standard Boolean operators (negation \lstinline|!|, conjunction \lstinline|&|, disjunction \lstinline'|', and implication \lstinline|=>|) together with a set of temporal and spatial modalities. 

Temporal properties are specified via the standard \lstinline|until| and \lstinline|since| operators (see e.g. \cite{mn13,MalerN04}) from which we can derive the future  \lstinline|eventually| and  \lstinline|globally| operators and the corresponding past variants,  \lstinline|once| and  \lstinline|historically|. All these operators may take an interval of the form \lstinline|[a,b]|, with \lstinline|a,b| $\in \mathbb{R}_{\geq 0}$, to define where the property should be  evaluated. The interval can be omitted in case of unbounded temporal operators.

Spatial modalities, instead, are \lstinline|reach| and  \lstinline|escape| operators, and the derivable operators \lstinline|somewhere| and  \lstinline|everywhere|. All these operators may be decorated with a \emph{distance interval}  \lstinline|[a,b]| and a \emph{distance expression}.
The distance expression consists of an expression that is used to compute the \emph{length} of an edge. If omitted, the real value \lstinline|1.0| is used. 
 \begin{figure}
    \centering

\begin{minipage}{37em}
\begin{lstlisting}[frame=single]
	(atomicExpression) 
	| ! Formula 
	| Formula & Formula 
	| Formula | Formula 
	| Formula => Formula 
	| Formula until [a b] Formula 
	| Formula since [a b] Formula  
	| eventually [a b] Formula 
	| globally [a b] Formula 
	| once [a b] Formula  
	| historically [a b] Formula 
	| escape(distanceExpression)[a b] Formula 
	| Formula reach (distanceExpression)[a b] Formula 
	| somewhere(distanceExpression) [a b] Formula 
	| everywhere (distanceExpression) [a b] Formula 
	| {Formula}
\end{lstlisting}
\end{minipage}
    \caption{STREL syntax.}
    \label{fig:syntax}
\end{figure}

To describe the spatial operators, we consider some examples.
Let us first consider the following property of the script:
 
\vspace{2ex}
\noindent \lstinline|P1 = (nodeType==3) reach(hop)[0, 1]{(nodeType==1)|$|$\lstinline|(nodeType==2)}|
\vspace{2ex}

\noindent \lstinline|P1| holds if from a node of type \lstinline|3| (an \emph{end device}), we can reach a node of type \lstinline|1| or \lstinline|2| (a \emph{coordinator} or a \emph{router}), following a path in the spatial graph such that the \lstinline|hop| distance along this path (i.e. its number of edges) is not bigger than 1. This property specifies that \emph{"end device should be directly connected to a router or the coordinator"}. 

The \lstinline|reach| operator allows us to express properties related to the existence of a path. The other operator of STREL,  \lstinline|escape|, can be used to express the ability of move away from a given point. 
Let us consider the following property:

\vspace{2ex}
\noindent \lstinline|P2 = escape(hop)[5,inf] (battery > 0.5) |
\vspace{2ex}

P2 states that from a given location, we can find a path of (hop) length at least 5 such that all nodes along the path have a battery level greater than 0.5, i.e. that a message will be forwarded along a connection with no risk of power failure.  

To specify properties \emph{around} a given location, operators \lstinline|somewhere| and \lstinline|everywhere| can be used.
For instance, we can consider the following property:

\vspace{2ex}
\noindent \lstinline|P3 = somewhere(dist)[0,250] (battery > 0.5)) |
\vspace{2ex}

\noindent
P3 is satisfied (at a given location) whenever there is a node at a distance between \lstinline|0| and \lstinline|250| having a \lstinline|battery| greater than \lstinline|0.5|. 
In this formula the distance is computed by summing the value \lstinline|dist| of traversed edges. 
The \lstinline|everywhere| operators works in a similar way, however it  requires that its subformula holds in all nodes satisfying the distance constraints.


Note that both \lstinline|reach| and \lstinline|escape| are existential operators, as they predicate the existence of a path with certain properties, and all the properties are interpreted at a given location, at a given time.
Temporal and spatial operators can be nested as for example:

\vspace{2ex}
\noindent \lstinline|PT1 = (battery <= 0.5)reach(hop)[0, 10] eventually(battery > 0.5) |
\vspace{2ex}

\noindent \lstinline|PT1| holds if each node can reach a node in less than 10 hops where the battery is greater than 0.5 in at least one time step in the next 5 time units.
We will show a second example later, but for more formal details and examples about STREL, we refer to~\cite{Bartocci17memocode} and the to tool documentation. 


\subsection{Using \textsc{MoonLight} in \textsc{Matlab} \texttrademark{}}
To use \textsc{MoonLight} in \textsc{Matlab} \texttrademark{} one has just to run the installation script (named \lstinline|install.m|) distributed with \textsc{MoonLight}. A detailed description of the installation process is available at the tool web site. After this installation, \textsc{MoonLight} becomes available to be used in the \textsc{Matlab} \texttrademark{} environment. 
In  Fig. \ref{fig:script2} a simple example is presented showing the main features of this module.

\begin{figure}
    \centering
\begin{lstlisting}
s = ScriptLoader.loadFromFile("sensorNetMonitorScript");
m = s.getMonitor("Ppar");
param = 5;
br = m.monitor(spatialModel,time,values,param);
script.setMinMaxDomain(); 
rr = m.monitor(spatialModel,time,values,param);
\end{lstlisting}
    \caption{\small Example of \textsc{Matlab}\texttrademark{} script that  uses \textsc{MoonLight}}
    \label{fig:script2}
    \vspace{-3ex}
\end{figure}


The function \lstinline|ScriptLoader.loadFromFile| loads the script.  It takes as parameter the file name containing the script to load (see line 1). 
After this operation is performed, a Java class is generated from the script and dynamically loaded. A reference to this object is returned to be used later. 
In the provided code, the script of Fig.~\ref{fig:script1} is loaded from the file \lstinline|sensorNetMonitorScript|.
%

After a script is loaded, the \emph{monitors} defined inside can be instantiated.  In the example of Fig.~\ref{fig:script2} the monitor associated with formula named \lstinline|Ppar| is retrieved (line 2). 
When we have the monitor we can use it to verify the satisfaction of the property over a given spatio-temporal signal. 
This is done by invoking the method \lstinline|monitor| of the just loaded monitor. 
This function takes the following parameters:
\begin{itemize}
    \item \lstinline|spatialModel|  is an array of \textsc{Matlab} \texttrademark\enskip graph structures specifying the spatial structure at each point in time; in the sensor network example, for each time step \texttt{i},
    \lstinline|spatialModel{i}.Edges| represents the adjacent list of the graph.
    This input is omitted when purely temporal models are considered.
    \item \lstinline|time|  is an array of time points at which observations are provided.
    \item \lstinline|values|  is a map (a cell array), with a cell for each node. In each cell, there is a matrix $n \times m$ where each row represents the values of the signals at the time points specified by \lstinline|time| (with $n$ equal to the number of time points and  $m$ the number of the considered signals); in the sensor network example, each node has a 3 signals representing the node's type,  battery, and  temperature. We represent different types of nodes using integer numbers, 1, 2, and 3 to represents \emph{coordinator}, \emph{router}, and \emph{end-device} respectively.
    This input is a simple matrix in case of purely temporal models.
    \item \lstinline|param|  is used to instantiate the parameter \lstinline|k| of formula \lstinline|Ppar|.
\end{itemize}

The output \texttt{br} from line 4 in Fig.~\ref{fig:script2} is similar to the input signal. It is a map that associates a Boolean-value (for the Boolean semantics)  or a real-value  signal (for the quantitative semantics) with each node, i.e. the Boolean or quantitative satisfaction at each time in each node. Finally, line 5 shows how to set the quantitative semantics ( in the Boolean case: \lstinline|moonlightScript.setBooleanDomain()|).

In Fig.~\ref{fig:spatialEval} (left), we can see the Boolean satisfaction at time zero of each node with respect the formula P1 of our script example in Fig~\ref{fig:script1}.  
The {\color{blue}blue nodes} (marked with a V) on the plot of Fig.~\ref{fig:spatialEval}(left) correspond to the nodes that satisfies the property, i.e. the end devices that reach a router or a coordinator with at most 1 hop. Fig.~\ref{fig:spatialEval} (right) shows the satisfaction of formula:

\vspace{3ex}
{\small \noindent \noindent \lstinline|P4=(nodeType==3) reach(hop)[0,1]{(nodeType==2) reach(hop)[0,5](nodeType==1)}|
}
\vspace{1ex}

\noindent P4 holds only in the nodes connected directly to the coordinator or to routers that can reach the coordinator through a maximum of four other routers. We can see that nodes 3, 4, 5 and 9 satisfy P1 but not \lstinline|P4|. Property \lstinline|PT2 = globally P4| can be used to check that \lstinline|P4| is true in each time step.

\section{Experimental Evaluation}

Our experiments were performed on a workstation with an Intel Core i7-5820K (6 cores, 3.30GHz) and 32GB RAM, running Linux Ubuntu 16.04.6 LTS, \textsc{Matlab} \texttrademark \enskip R2020a and \textsc{OpenJDK} 64-Bit Server VM 1.8.0\_252.

\subsection{\bf Temporal evaluation: monitoring Signal Temporal Logic}
We consider the \emph{Automatic Transmission} example in~\cite{HoxhaAF14}.  This benchmark consists of a \textsc{Matlab} \texttrademark/Simulink deterministic model of an automatic transmission controller.
The model has two inputs (the throttle and the break) and two outputs:
the speed of the engine $\omega$ (RPM) and the speed of the vehicle $v$ (mph). We monitor the robustness of  four requirements in~\cite{HoxhaAF14}:
\vspace{1ex}

\noindent {\color{red} (R1)}
 The engine speed never reaches $\bar{\omega}$: 
 \footnotesize{{\lstinline|globally| }$(\omega < \bar{\omega}$)}\normalsize

\noindent {\color{blue} (R2)} The engine and the vehicle speed never reaches $\bar{\omega}$ and $\bar{v}$ resp.:\\ \footnotesize{{\lstinline|globally| } $((\omega < \bar{\omega})$ \& $(v < \bar{v})$)}\normalsize
    
\noindent {\color{orange} (R3)} If engine speed is always less than $\bar{\omega}$, then vehicle speed can not exceed $\bar{v}$ in less than $T$ sec.:
{\footnotesize{ \lstinline|!(eventually [0, T] | $(v > \bar{v})$ \& \lstinline|globally| $(\omega < \bar{\omega}))$}}\normalsize

     \noindent {\color{brown} (R4)} Within T sec. the vehicle speed is above $\bar{v}$ and from that point on the engine speed is always less than $\bar{\omega}$:  {\footnotesize{\lstinline|eventually [0,T]| $((v \geq \bar{v})$ \& \lstinline|globally| $(\omega < \bar{\omega}))$}}\normalsize
\vspace{1ex}

\begin{figure}[h]
\includegraphics[width=\linewidth]{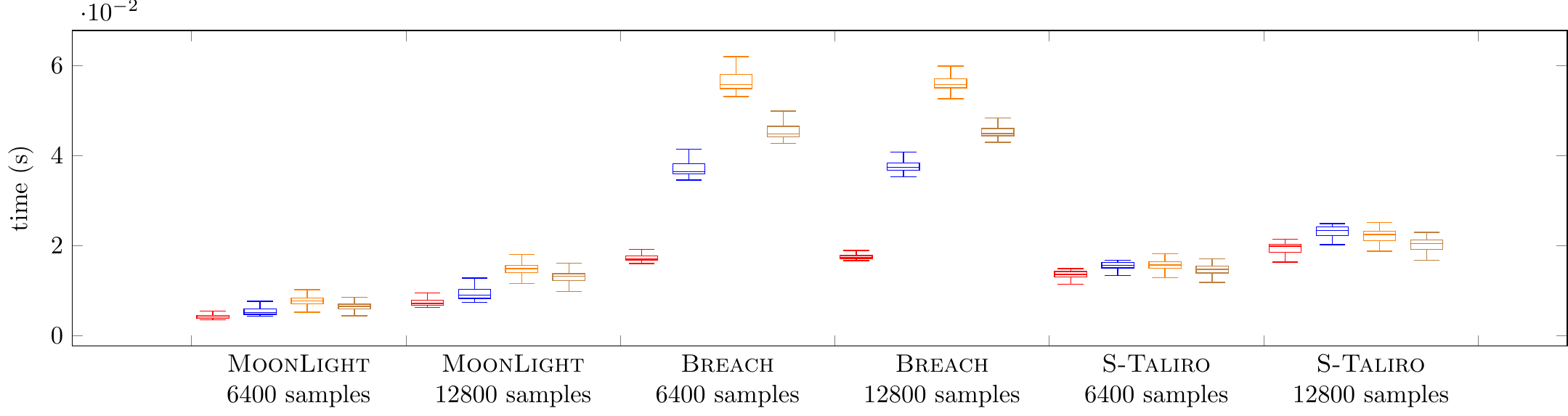}
\caption{The comparison of the computational time (in sec.) between \textsc{MoonLight}, \textsc{Breach} and \textsc{S-Taliro} for simulation traces with different length. The different colors represent the result for different requirements: {\color{red} (R1)}, {\color{blue} (R2)}, {\color{orange} (R3)} and {\color{brown} (R4)}.} 
	\label{fig:temporalEval}
\end{figure}

 We randomly generated 20 different input traces with 6400 samples and another 20 with 12800 samples (0.01 sec. of sampling time). For each input trace, we simulated the model and we  monitored the robustness of the four requirements over the outputs by varying the parameters $\bar{v} \in \{120, 160, 170, 200\}$, $\bar{w} \in \{4500,5000,5200,5500\}$ and $T \in \{4,8,10,20\}$. For a fixed combination of parameters and output traces, we repeated the monitoring experiment 20 times and we considered the mean of the execution times.
 In Figure~\ref{fig:temporalEval}, we 
compare the performance of our \textsc{Moonlight} monitors with \textsc{S-Taliro}~\cite{staliro} and \textsc{Breach}~\cite{breach} using
bloxplots representing the quartiles of the execution times distribution for monitoring each requirement with each tool. The graph shows a good performance
of \textsc{Moonlight} with respect to the other tools. However, it is important to note that 
\textsc{Breach} considers 
piece-wise linear signals 
and computes the
interpolation between two 
consecutive samples when necessary, while our tool and \textsc{S-Taliro} interpret the signal step-wise.




\lstset{keywordstyle=\color{black}}

\subsection{\bf Spatio-Temporal Evaluation}
We evaluate the scalability of the spatial operators in the running example varying the number of nodes of the graph: $N = 10, 100, 1000$. Note that the number of edges is around $1700$ for $N=100$ and $8000$ for $N=1000$. We monitor the Boolean (B) and the quantitative (Q) semantics of the requirements presented in Section \ref{sec:moonlight_script}, excluding P1.
For the spatio-temporal requirements we consider K time step, for $K=10,100$. 
We repeat the monitoring experiments 50 times for each N. %
Table \ref{tab:spatialEval} shows the average execution time. For spatial formulas, we can see that formula (P4) performs better than the other two. (P2) is slower because monitoring algorithms of the reach and somewhere are $O(n^2)$ i.e. linear in the number of edges and quadratic in the number of vertexes, while the one for escape is $O(n^3)$. As expected, the Boolean semantics is faster than the quantitative one: it can reach sooner the fixed point. Formula (P3) is the slowest due to the fact that uses the euclidean distance while formula (P2) and (P4) the lighter hop distance. 
 For spatio-temporal formulas, the reason why (PT1) is much faster than (PT2) is that (PT1) has a temporal subformula, hence the number of time steps can be dramatically reduced before monitoring the spatial part. This does not happen for (PT2), where the operators are inverted. In this case the difference between the two semantics is more evident. 
For static graphs and properties restricted to everywhere and somewhere spatial modalities, the performances are similar to \textsc{jSSTL}~\cite{NenziBL16,NenziBCLM18}. Further experiments can be found in the tool release.

\begin{table}[]
\begin{center}
\scalebox{0.9}{
\begin{tabular}{|c|c|c|c|c|c|c|c|c|c|c|c|c|c|c|}
\hline 
 & \multicolumn{6}{|c|}{K = 1} & \multicolumn{4}{c|}{K = 10} & \multicolumn{4}{c|}{K = 100} \\\hline
       &  \multicolumn{2}{c|}{P2} & \multicolumn{2}{c|}{P3} & \multicolumn{2}{c|}{P4}& \multicolumn{2}{c|}{PT1} &\multicolumn{2}{c|}{PT2} & \multicolumn{2}{c|}{PT1} & \multicolumn{2}{c|}{PT2} \\ \hline
  N   & B           & Q         & B         & Q         & B        & Q        & B       & Q      & B      & Q      & B       & Q     & B      & Q      \\ \hline
10    &   $0.0031$  &  $0.0032$ &  $0.0031$ & $0.0029$  & $0.0027$ & $0.0026$ & $0.021$ & $0.021$& $0.026$& $0.021$& $0.0.24$& $0.17$& $0.17$ & $0.17$ \\ \hline
100   &   $0.013$   &  $0.020$  &  $0.042$  & $0.0419$   & $0.0088$  & $0.0084$  & $0.067$ & $0.081$& $0.10$ & $0.14$ & $0.76$  & $0.73$& $1.02$ & $1.5$ \\ \hline
1000  &   $0.86$    &  $4.97$   &  $16.91$   & $16.95$    & $0.11$  & $0.12$   & $0.60$  & $0.76$ & $6.18$ & $14.74$& $6.68$  & $7.29$& $99.17$& $276.8$\\ \hline
\end{tabular}
}
\end{center}
\caption{\small The comparison of the computational time (in sec) with respect the number of nodes of the graph N for formulas (P2), (P3), (P4),  and with respect N and the number of time steps K for formulas (PT1), and (PT2) for Boolean (B) and quantitative (Q) semantics.}
\label{tab:spatialEval}
  \vspace{-2ex}
\end{table}

\section{Conclusion}
\textsc{MoonLight} provides a lightweight and very flexible monitoring tool for temporal and spatio-temporal properties of mobile and spatially arranged CPS. The possibility to use a dedicated \textsc{Matlab} \texttrademark \enskip interface enables to easily integrate \textsc{MoonLight} as a component in other tool chains implementing more sophisticated computer-aided verification and synthesis techniques such as falsification analysis and parameter synthesis. In the near future, we plan to add also a Python interface and to extend the tool with new functionalities such as the support parallelized and online monitoring.

\bibliographystyle{splncs04}
\bibliography{refs}

\end{document}